\documentclass[%
 reprint,
superscriptaddress,
 amsmath,amssymb,
 aps,
prl,
]{revtex4-2}

\usepackage{graphicx}
\usepackage{dcolumn}
\usepackage{bm}
\usepackage{comment}
\usepackage{multirow}
\usepackage{color}
\usepackage{ulem}

\usepackage{hyperref}

\usepackage{amsthm}
\theoremstyle{plain}

\newtheorem*{thm*}{Main Theorem}

\newtheorem*{lem*}{Lemma}
\theoremstyle{definition}

\newtheorem*{dfn*}{Definition}

\newcommand{\bmat}[1]{\begin{matrix} #1 \end{matrix}}
\newcommand{\loq}[1]{\begin{array}{l} (#1) \end{array}}

\newcommand{\eq}[1]{\begin{eqnarray} #1 \end{eqnarray}}
\newcommand{\nx}{ \nonumber \\}

\def \({\left(}
\def \){\right)}
\def \[{\left[}
\def \]{\right]}

\newcommand{\sumi}{\sum_{i=1}^N}

\newcommand{\jx}{J_X X_i X_{i+1}}
\newcommand{\jy}{J_Y Y_i Y_{i+1}}
\newcommand{\jz}{J_Z Z_i Z_{i+1}}
\newcommand{\hx}{h_X X_i}
\newcommand{\hy}{h_Y Y_i}
\newcommand{\hz}{h_Z Z_i}

\newcommand{\boldA}{\boldsymbol{A}}
\newcommand{\boldB}{\boldsymbol{B}}

\newcommand{\ak}{{\boldA}^k}
\newcommand{\aki}{\ak_i}
\newcommand{\qa}{q_{\aki}}

\begin{document}

\preprint{APS/123-QED}

\title{Complete Classification of Integrability and Non-integrability\\ for Spin-1/2 Chain with Symmetric Nearest-Neighbor Interaction}
\author{Mizuki Yamaguchi}
\email{yamaguchi-q@g.ecc.u-tokyo.ac.jp}
\affiliation{
Graduate School of Arts and Sciences, The University of Tokyo,
3-8-1 Komaba, Meguro, Tokyo 153-8902, Japan}
\author{Yuuya Chiba}
\email{yuya.chiba@riken.jp}
\affiliation{Nonequilibrium Quantum Statistical Mechanics RIKEN Hakubi Research Team,
RIKEN Cluster for Pioneering Research (CPR), 2-1 Hirosawa, Wako, Saitama 351-0198, Japan}

\author{Naoto Shiraishi}
\email{shiraishi@phys.c.u-tokyo.ac.jp}
\affiliation{
Graduate School of Arts and Sciences, The University of Tokyo,
3-8-1 Komaba, Meguro, Tokyo 153-8902, Japan}

\begin{abstract}
General spin-1/2 chains with symmetric nearest-neighbor interaction are studied. We rigorously prove that all spin models in this class, except for known integrable systems, are non-integrable in the sense that they possess no nontrivial local conserved quantities. 
This result confirms that there are no missing integrable systems, i.e., integrable systems in this class are exactly those that are already known.
In addition, this result excludes the possibility of intermediate systems which have a finite number of nontrivial local conserved quantities.
Our findings support the expectation that integrable systems are exceptional in quantum many-body systems and most systems are non-integrable.

\end{abstract}
\maketitle


\paragraph*{Introduction---}\label{par:Introduction}

In the study of quantum many-body systems, distinction between integrable and non-integrable systems is crucial. Quantum integrable systems have Hamiltonians whose eigenvalues and eigenstates are exactly solvable \cite{baxter2016exactly, jimbo1994algebraic, faddeev1996algebraic, takahashi1999thermodynamics}. Integrable systems have been studied for almost a century, and various models have been unveiled to be integrable \cite{1931.Bethe.ZP.71, lieb1961two, yang1966one, lieb1968absence, baxter1971one, maassarani1998xxc, kitaev2001unpaired, yanagihara2020exact, sutherland1975model, Takhtajan:1982jeo, babujian1982exact, barber1989spectrum}.
Integrable systems are highly useful in that many important physical quantities can be calculated exactly. 
The solvability of integrable systems is thought to be a consequence of the presence of an infinite number of local conserved quantities \cite{caux2011remarks, gogolin2016equilibration}. Here, a local conserved quantity is a conserved quantity given by a sum of operators acting on a few spatially consecutive sites. To the best of our knowledge, all known integrable systems with local and shift-invariant Hamiltonians discovered to date possess infinite local conserved quantities \cite{grabowski1995structure, fagotti2013reduced}.

On the other hand, unlike conventional macroscopic systems, integrable systems exhibit some anomalous behaviors. For instance, thermalization of isolated systems, a fundamental principle of thermodynamics, does not occur in integrable systems due to the obstruction of local conserved quantities \cite{rigol2007relaxation, pozsgay2013generalized, vidmar2016generalized}, as observed in experiments using ultra-cold atoms \cite{kinoshita2006quantum, langen2015experimental, reiter2021engineering}. In addition, other empirical laws including the Kubo formula in linear response theory and Fourier's law in thermal conduction do not apply to integrable systems \cite{mazur1969non, suzuki1971ergodicity, zotos1997transport, saito2003strong}. Since these laws are empirically confirmed, it is believed that integrable systems are exceptional and almost all systems are non-integrable in the sense of the absence of nontrivial conserved quantities \cite{mori2018thermalization, shiraishi2019proof}.
Various numerical experiments support this belief. For instance, the distribution of the energy level spacings is qualitatively different between integrable and non-integrable systems, and generic systems follow the distribution for the latter one (the Wigner-Dyson distribution) \cite{dyson1962statistical, wigner1993characteristic, rigol2009breakdown, santos2010onset, atas2013distribution}.

In spite of these strong expectations, ubiquitousness of non-integrability has not yet been addressed from an analytical perspective. 
In contrast to vast literature on integrable systems, it has been considered to be difficult to prove non-integrability. In fact, a rigorous proof of non-integrability was elusive for a long time.

Recently, non-integrability was rigorously proved in the XYZ model with Z magnetic field \cite{shiraishi2019proof}.
Triggered by this work, several models, the mixed-field Ising model \cite{chiba2024proof}, the PXP model \cite{park2024proof}, and the Heisenberg model with next-nearest-neighbor interaction~\cite{shiraishi2024absence}, were proved to be non-integrable. However, these studies treat non-integrability of individual systems. To substantiate the expectation that most systems are inherently non-integrable, it is necessary to prove non-integrability of a broad class of systems.

In this paper we demonstrate the ubiquitousness of non-integrability within a general class of systems. Precisely, we consider general spin-1/2 chains with shift-invariant and parity-symmetric nearest-neighbor interactions.
We rigorously prove that all systems except known integrable systems are indeed non-integrable, i.e., they have no nontrivial local conserved quantity. This class of models in consideration includes many well-established integrable systems, such as the Heisenberg model and the transverse-field Ising model. Our result establishes that there is no further integrable system beyond known ones. In addition, our result also shows that any system has infinitely many local conserved quantities or no nontrivial local conserved quantity, and there is no intermediate system with a finite number of nontrivial local conserved quantities. This result supports the view that integrable systems are exceptional and almost all systems are non-integrable.

\paragraph*{Setup and main result---}\label{par:Setup}

We consider a general spin-1/2 chain with $N$ spins satisfying the following conditions:
 the Hamiltonian is translationally invariant, and the Hamiltonian  consists only of parity symmetric nearest-neighbor interaction terms and magnetic field terms.
Precisely, the Hamiltonian in consideration is given by
\eq{
H= \sumi \sum_{\alpha, \beta\in \{ x,y,z\}} J_{\alpha\beta}\sigma_i^\alpha \sigma_{i+1}^\beta +\sum_{\alpha\in \{ x,y,z\}} h_\alpha \sigma_i^\alpha \label{general}
}
with $J_{\alpha\beta}=J_{\beta\alpha}$ for any $\alpha$ and $\beta$.
We identify site $N+1$ to site $1$, meaning the periodic boundary condition.
We abbreviate Pauli matrices $\sigma_i^x$, $\sigma_i^y$, and $\sigma_i^z$ as $X_i$, $Y_i$, and $Z_i$, respectively.

In this Letter, we use the term {\it non-integrable} to signify that the system has no nontrivial local conserved quantity.
Here, we say that a quantity is {\it conserved} if it commutes with the Hamiltonian of the system, and a quantity is {\it local} if it can be written by a sum of operators which act on $O(1)$ consecutive sites.
In particular, a quantity that can be written by a sum of operators on $k$ consecutive sites is called a {\it $k$-support} quantity.
Notice that locality means spatial locality of the range of interactions, not few-body interactions.
For instance, $\sumi X_i Y_{i+4}$ is a 5-support local quantity (not 2-support), and $\sumi Z_i Z_{i+N/2}$ is {\it not} a local quantity.
A local conserved quantities with at least 3-support is referred to as nontrivial.
If a system has an infinite number of local conserved quantities in the thermodynamic limit, we call this system {\it integrable}, and if it has a finite number of nontrivial local conserved quantity, we call this system {\it partially integrable} in correspondence with the terminology in studies of classical Hamiltonian systems \cite{goriely1996integrability}.

\begin{figure}[t]
\includegraphics[width=6cm]{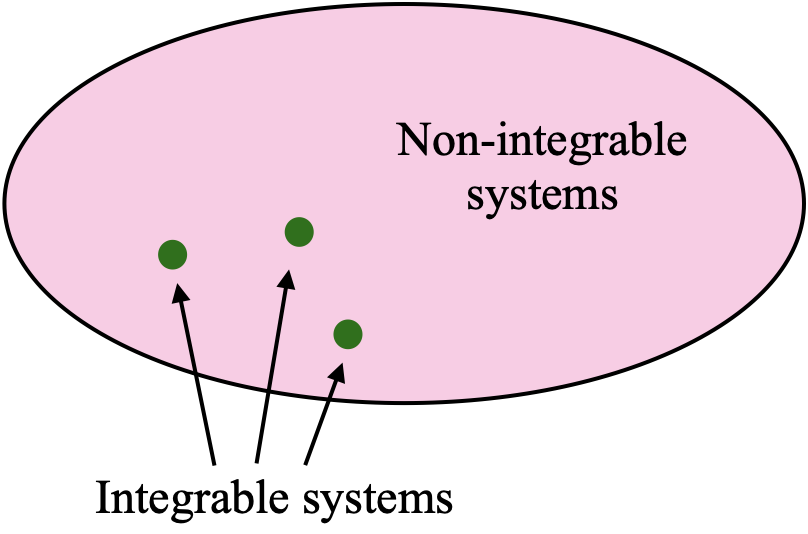}
\caption{A schematic of what we have proven in this Letter. All systems except for known integrable systems are rigorously proven to be non-integrable, which dominates the space of possible models in this class.}
\end{figure}

Below we complete the classification of the systems represented by Eq.~\eqref{general} with a rigorous proof from the viewpoint of integrability.
We prove that all systems represented by Eq.~\eqref{general} are classified into the following two groups:
\begin{itemize}
    \item \underline{Integrable systems}: The XXX (Heisenberg) model \cite{1931.Bethe.ZP.71}, the XXZ model \cite{yang1966one}, the XYZ model \cite{baxter1971one}, the XY model \cite{lieb1961two}, the transverse-field Ising model \cite{lieb1961two}, trivial classical models, and their global spin rotations.
    \item \underline{Non-integrable systems}: All systems except the above.
\end{itemize}
This is our main result of this Letter.
Since listed models as integrable systems have already been shown to be integrable in previous literature, what we need to prove is the non-integrability of all the remaining systems.
We present the proof idea and outline the proof in this Letter. Its complete proof is provided in the companion paper \cite{FULLPAPER}.

\paragraph*{Implications of main result---}
Before going to the proof idea and the proof outline, we here highlight the meaning of our result.
Our main result manifests the fact that there is no missing integrable system which is awaited to be discovered.
By conventional approaches in the field of integrable systems, it is almost hopeless to prove that obtained integrable models are the complete list of integrable models and there is no oversight.
Our non-integrability proof indeed does this, and proves that there is no more integrable systems in the form of Eq.~\eqref{general}.

This result also supports the expectation that integrable systems are highly exceptional systems and most generic systems are non-integrable.
Since non-integrability is necessary to fulfill the linear response theory and thermalization, the ubiquitousness of non-integrability is strongly believed.
However, this belief is usually confirmed only by numerical simulations, and its rigorous mathematical foundation has been elusive.
Our result serves as the first mathematical foundation of the above belief.

In addition, our result demonstrates the absence of a partially integrable system in this class, i.e., an intermediate system with finite nontrivial local conserved quantities.
Although all known generic models have sufficiently many local conserved quantities or no nontrivial local conserved quantity, in principle there may exist an intermediate system which has a few nontrivial local conserved quantities.
This is an open problem in the research field of integrable systems \cite{grabowski1995integrability}. Our result solves this problem in negative as long as the system is described in the form of Eq.~\eqref{general}.

\begin{table*}[t!]
\def\arraystretch{1.5}
\caption{We classify integrability/non-integrability of all systems expressed in Eq.~\eqref{standard}. 
Any system in the form of Eq.~\eqref{general} is reduced to Eq.~\eqref{standard}.
Mark $*$ in the table means that similar classifications apply to cases where Pauli $X$, $Y$, and $Z$ are permuted.
}
\begin{tabular}{|c|c|c|c|c|}
\hline
Rank & Standard form ($J_X,J_Y,J_Z \neq 0$) & Detailed conditions & Integrability & Name \& Reference \\ \hline
$0$ & $ \sumi \hz$ & - & Integrable & Free spins \\ \hline
\multirow{3}{*}{$1$} & \multirow{3}{*}{$\displaystyle \sumi ( \jz + \hx + \hz)$} & $h_X = 0$ & Integrable & Longitudinal-field Ising model \\ \cline{3-5}
 & & $h_X \neq 0, h_Z = 0$ & Integrable & Transverse-field Ising model \cite{lieb1961two} \\ \cline{3-5}
 & & $h_X \neq 0, h_Z \neq 0$ & Non-integrable & Proven in \cite{chiba2024proof} \\ \hline
\multirow{2}{*}{$2$} & \multirow{2}{*}{\begin{tabular}{c}$\displaystyle \sumi \loq{ \jx + \jy \\  + \hx + \hy + \hz}$ \end{tabular}} & $(h_X,h_Y) = (0,0)$ & Integrable & XY model \cite{lieb1961two} \\ \cline{3-5}
& & $(h_X,h_Y) \neq (0,0)$ & Non-integrable & \bf{Proven in this Letter} \\ \hline
\multirow{9}{*}{$3$} & \multirow{9}{*}{\begin{tabular}{c}$\displaystyle \sumi \loq{ \jx + \jy + \jz \\  + \hx + \hy + \hz}$ \end{tabular}} & $J_X = J_Y = J_Z$ & Integrable & XXX model \cite{1931.Bethe.ZP.71} \\ \cline{3-5}
& & \begin{tabular}{c}*\quad$J_X=J_Y \neq J_Z,$\quad{} \\ $(h_X,h_Y) = (0,0)$\end{tabular} & Integrable & XXZ model \cite{yang1966one} \\ \cline{3-5}
& & \begin{tabular}{c}*\quad$J_X=J_Y \neq J_Z,$\quad{} \\ $(h_X,h_Y) \neq (0,0)$\end{tabular} & Non-integrable & \bf{Proven in this Letter} \\ \cline{3-5}
& & \begin{tabular}{c}$J_X \neq J_Y, J_Y \neq J_Z, J_Z \neq J_X,$\\$(h_X,h_Y,h_Z) = (0,0,0)$\end{tabular} & Integrable & XYZ model \cite{baxter1971one} \\ \cline{3-5}
& & \begin{tabular}{c}$J_X \neq J_Y, J_Y \neq J_Z, J_Z \neq J_X,$\\$(h_X,h_Y,h_Z) \neq (0,0,0)$\end{tabular} & Non-integrable & \begin{tabular}{c} \bf{Proven in this Letter} \\ (Special case: \cite{shiraishi2019proof}) \end{tabular} \\ \hline
\end{tabular}
\label{tab:classification}
\end{table*}

\paragraph*{Proof idea---}
In accordance with Refs.~\cite{shiraishi2019proof, chiba2024proof, park2024proof, shiraishi2024absence}, we prove the absence of local conserved quantities by the following approach:
We first expand a candidate of a local conserved quantity by strings of Pauli operators and the identity ($X,Y,Z,$ and $I$).
We call these strings $l$-support {\it Pauli strings} $\boldA_i^l = A_i A_{i+1} \cdots A_{i+l-1}$, where $A_j\in \{X,Y,Z,I\}$ ($j\neq i, i+l-1$) and $A_i, A_{i+l-1}\in \{ X,Y,Z\}$ are satisfied.
Since the basis of $k$-support quantities is $l(\leq k)$-support Pauli strings (hereafter simply called $l$-support operators),
any $k$-support quantity can be expanded as
\eq{
Q = \sum_{l=0}^k \sum_{i=1}^N \sum_{\boldA^l} q_{\boldA^l_i} \boldA^l_i ~,
\label{input}
}
where all coefficients $\{ q_{A_i^l} \}$ are real numbers.
Moreover, since $H$ is a 2-support quantity, the commutator of $Q$ and $H$ is a at-most-$(k+1)$-support quantity, which can be expanded as 
\eq{
\frac{1}{2i} [Q, H] = \sum_{l=0}^{k+1} \sum_{i=1}^N \sum_{\boldB^l} r_{\boldB^l_i} \boldB^l_i ~.
\label{output}
}

The conservation condition, $r_{\boldB^l_i} = 0$ for all $\boldB^l_i$, yields a system of linear equations of $\{q_{\boldA^l_i}\}$. The goal of the proof of non-integrability is to show that these equations do not have solutions except for $\qa = 0$ for all $\aki$, which means that $Q$ cannot be a $k$-support conserved quantity. Since we need to treat general $k$, we should find a systematic way to show $\qa = 0$ for all $\aki$.

We first simplify the Hamiltonian \eqref{general} by diagonalizing the interaction matrix $(J_{\alpha\beta})$ with the global spin rotation $\begin{pmatrix}X' \\ Y' \\ Z'\end{pmatrix} = R \begin{pmatrix}X \\ Y \\ Z\end{pmatrix}$.
Through this, a general Hamiltonian \eqref{general} can be transformed into the following standard form:
\eq{
H = \sumi \loq{  \jx  +\jy +\jz \\ +\hx +\hy +\hz } ~.
\label{standard}
}
Notice that the integrability/non-integrability does not change through this transformation.
Hence, it suffices to treat only the standard form \eqref{standard}.
Our complete classification for the standard form is summarized in Table~\ref{tab:classification}.

Note that the procedure for the non-integrability proof differs depending on the rank of $(J_{\alpha\beta})$, i.e., the number of non-zero elements in $J_X, J_Y,$ and $J_Z$, because linear equations of $\{q_{{\boldA}^l_i}\}$ have contributions on only terms with nonzero interaction coefficients.
Below, we prove the absence of 3-support local conserved quantity in the case of rank 2 (i.e., $J_X,J_Y \neq 0$ and $J_Z=0$) as an example
(See the companion paper \cite{FULLPAPER} for the full proof).
Our proof for this case consists of two steps.
In Step 1 we use the condition $\{ r_{\boldB_i^{k+1}} = 0\}$ to eliminate most of coefficients $\{q_{\boldA_i^k}\}$.
In Step 2 we show that the remaining coefficients are also zero by using the condition $\{ r_{\boldB_i^k} = 0\}$.

\paragraph*{Proof for 3-support conserved quantities with rank 2: Step 1---}
To consider linear equations of $\{\qa\}$ given by the conservation condition, we introduce a useful expression called  {\it column expression}, with which we can grasp commutation relations visually.
Let us take as an example a commutator between $\boldA^3_i =Z_iY_{i+1}X_{i+2}$ in $Q$ and $Y_{i+2}Y_{i+3}$ in $H$ generates $\boldB^4_i = Z_iY_{i+1}Z_{i+2}Y_{i+3}$ as
\eq{ [Z_i Y_{i+1} X_{i+2}, Y_{i+2} Y_{i+3}] = +2i \, Z_i Y_{i+1} Z_{i+2} Y_{i+3}.}
In the column expression, this commutation relation is expressed as
\eq{
\bmat{
 & Z_i & Y_{i+1} & X_{i+2} \\
 &     &         & Y_{i+2} & Y_{i+3} \\ \hline
+& Z_i & Y_{i+1} & Z_{i+2} & Y_{i+3}
}~.
\label{ZYZY}
}
Here, the horizontal line represents commutation operation, and the horizontal positions of three operators reflect the spatial positions of the sites they act on.
The coefficient $2i$ is dropped.

Since the coefficient of a 4-support operator $\boldB_i^4$ in $[Q,H]$ is zero due to the conservation condition, each 4-support operator $\boldB_i^4$ accompanies a linear equation of $\{q_{\boldA_i^3}\}$ derived from the commutation relation.
For example, $\boldB^4_i = Z_iY_{i+1}Z_{i+2}Y_{i+3}$ is generated only by Eq.~\eqref{ZYZY} and no other commutator generates $Z_iY_{i+1}Z_{i+2}Y_{i+3}$, since the Hamiltonian does not have $ZZ$ interaction and thus we do not have the following form of commutator:
\eq{
\bmat{
 & & ? & ? & ? \\
 & ? &  ? &  &  \\ \hline
+& Z_i & Y_{i+1} & Z_{i+2} & Y_{i+3}
}~.
}
The above fact directly means
\eq{
J_Y q_{Z_iY_{i+1}X_{i+2}} = r_{Z_iY_{i+1}Z_{i+2}Y_{i+3}} =0~.
}
By similar arguments, we can show that $q_{Z_i A_{i+1} A_{i+2}}=0$ for arbitrary $A_{i+1}$ and $A_{i+2}$.
Due to the inversion symmetry, Pauli strings with the reverse order of above also have zero coefficients: $q_{A_i A_{i+1} Z_{i+2}} = 0$ for arbitrary $A_i$ and $A_{i+1}$.

We also observe that
\eq{
\bmat{
 & X_i & X & Y \\
 &     &         & X & X \\ \hline
-& X_i & X & Z & X
}
}
is the only contribution to $\boldB^4_i = X_i X_{i+1} Z_{i+2} X_{i+3}$, since we do not have the following form of commutator:
\eq{
\bmat{
 & & ? & ? & ? \\
 & X &  X &  &  \\ \hline
-& X_i & X & Z & X
}~.
}
This fact directly implies $q_{X_i X_{i+1} Y_{i+2}}=0$.
Moreover, by observing that
\eq{
\bmat{
 & X_i & I & Z \\
 &     &   & X & X \\ \hline
+& X_i & I & Y & X
}
}
is the only contribution to $\boldB^4_i = X_i I_{i+1} Z_{i+2} X_{i+3}$ and considering similar commutators, we find $q_{X_i X_{i+1} A_{i+2}} = q_{X_i I_{i+1} A_{i+2}} = q_{Y_i Y_{i+1} A_{i+2}} = q_{Y_i I_{i+1} A_{i+2}} = 0$ for arbitrary $A_{i+2}$.

Moreover, all contributions to $\boldB^4_i = X_iY_{i+1}Z_{i+2}X_{i+3}$ are listed as
\eq{
\bmat{
 & X_i & Y & Y \\
 &     &   & X & X \\ \hline
-& X_i & Y & Z & X
}\qquad
\bmat{
 &     & Z_{i+1} & Y & X \\
 & X & X & & \\ \hline
+& X_i & Y & Y & X
}~,
}
which lead to
\eq{
J_X q_{X_i Y_{i+1} Y_{i+2}} = J_X q_{Z_{i+1} Y_{i+2} X_{i+3}}~.
}
Recalling that $q_{Z_{i+1} Y_{i+2} X_{i+3}}$ has already been shown to be zero, we conclude that $q_{X_i Y_{i+1} Y_{i+1}}=0$.
In a similar manner to above, we can show $q_{X_i Y_{i+1} A_{i+2}} = 0$ as well as $q_{Y_i X_{i+1} A_{i+2}} = 0$ for arbitrary $A_{i+2}$.

In summary, $q_{\boldA_i^3} = 0$ holds for all $\boldA^3$ except for $\boldA^3 = XZX, XZY, YZX,$ and $YZY$.
In addition, we can show that the remaining coefficients satisfy the following linear relations:
\eq{
&\frac{q_{Y_i Z_{i+1} Y_{i+2}}}{J_Y} = \frac{q_{X_{i+1} Z_{i+2} X_{i+3}}}{J_X} = \frac{q_{Y_{i+2} Z_{i+3} Y_{i+4}}}{J_Y} = \cdots ~, \nx
&q_{Y_i Z_{i+1} X_{i+2}} = -q_{X_{i+1} Z_{i+2} Y_{i+3}} = q_{Y_{i+2} Z_{i+3} X_{i+4}} = \cdots ~.\nx
\label{linear}
}
Hence, to prove the absence of 3-support conserved quantity, it suffices to show the coefficients of $Y_1Z_2Y_3$, $Y_2Z_3Y_4$, $Y_1Z_2X_3$, and $Y_2Z_3X_4$ as zero.

\paragraph*{Proof for 3-support conserved quantities with rank 2: Step 2---}
We next focus on 3-support operators $\{ \boldB^3_i \}$ in $[Q,H]$ and derive $q_{X_iZ_{i+1}Y_{i+2}} = 0$ and $q_{X_iZ_{i+1}X_{i+2}} = 0$ for general $i$, which include the case of $i=1$ and 2.
Applying a global spin rotation if necessary, without loss of generality we assume $h_X \neq 0$.

Dropping all the contributions from $\boldA_i^3$ whose coefficients are shown to be zero in Step 1, we find that $\boldB_i^3 = Y_i Y_{i+1} Y_{i+2}$ is generated only by
\eq{
\bmat{
 & Y_i & Z & Y  \\
 &     & X &  \\ \hline
+& Y_i & Y & Y
}~,
}
which directly implies $h_X q_{Y_i Z_{i+1} Y_{i+2}} = 0$, meaning
\eq{
q_{Y_iZ_{i+1}Y_{i+2}}=0.
}
Inserting Eq.~\eqref{linear} to this, we have
\eq{
q_{X_iZ_{i+1}X_{i+2}}=0~.\label{step2_1}
}

Next, all commutators generating $\boldB_i^3 = Z_i Z_{i+1} X_{i+2}$ are
\eq{
\bmat{
 & Y_i & Z & X \\
 & X \\ \hline
-& Z_i & Z & X
}\qquad
\bmat{
 & X_i & Z & X \\
 & Y \\ \hline
+& Z_i & Z & X
}\qquad
\bmat{
 & Z_i & Y \\
 &     & X & X \\ \hline
-& Z_i & Z & X
}~,
}
and all commutators generating $\boldB_{i-1}^3 = X_{i-1} Y_i Y_{i+1}$ are
\eq{
\bmat{
 & X_{i-1} & Z & Y  \\
 &         & X &  \\ \hline
+& X_{i-1} & Y & Y
}\qquad
\bmat{
 &         & Z_i & Y \\
 & X & X \\ \hline
+& X_{i-1} & Y & Y
}~,
}
which lead to
\eq{
-h_X q_{Y_i Z_{i+1} X_{i+2}} + h_Y q_{X_i Z_{i+1} X_{i+2}} - J_X q_{Z_i Y_{i+1}} &=& 0 ~,\nx
h_X q_{X_{i-1} Z_i Y_{i+2}} + J_X q_{Z_i Y_{i+1}} &=& 0~.\nx
}
Summing these two relations and inserting Eq.~\eqref{linear} into it, we have
\eq{
-2h_X q_{Y_i Z_{i+1} X_{i+2}} + h_Y q_{X_i Z_{i+1} X_{i+2}} = 0~.
\label{step2_2}
}
Plugging Eq.~\eqref{step2_1} into Eq.~\eqref{step2_2}, we conclude that $q_{Y_i Z_{i+1} X_{i+2}}=0$.
This completes the proof of the absence of 3-support conserved quantities in rank 2 case.

\paragraph*{Outline of the proof for the general case---}
We here outline the proof of our main result.
The complete proof is given in the companion paper \cite{FULLPAPER}.

The proof of the absence of $k$-support conserved quantities in rank 2 case is quite close to that for $k=3$.
In Step 1, we employ the condition $\{ r_{\boldB^{k+1}_i}=0\}$ and show that (i) $q_{\boldA^k_i}=0$ holds except for $\boldA =XZZ\cdots ZZX$, $XZZ\cdots ZZY$, $YZZ\cdots ZZX$, and $YZZ \cdots ZZY$ and (ii) some linear relations hold between the remaining coefficients.
In Step 2, we employ the condition $\{ r_{\boldB^{k}_i}=0\}$ and show that all the remaining coefficients are zero.

In rank 1 case \cite{chiba2024proof} and rank 3 case \cite{shiraishi2019proof, FULLPAPER}, the framework of showing $q_{\boldA^k_i}=0$ for all $\boldA^k$ by first focusing on $\{ r_{\boldB^{k+1}_i}=0\}$ and then on $\{ r_{\boldB^{k}_i}=0\}$ is common to rank 2 case, while how to eliminate the coefficients $q_{\boldA^k_i}$ are quite different.

\paragraph*{Conclusion and discussion---}

We have performed a complete classification of integrability and non-integrability for spin-$1/2$ chain with symmetric nearest-neighbor interaction, based on the number of local conserved quantities.
All systems, other than known integrable systems, are rigorously proved to be non-integrable.
Precisely, it is established that systems are either integrable systems with infinite local conserved quantities or non-integrable systems with no local conserved quantity, and there is no intermediate system with a finite number of local conserved quantities.
Our result offers a rigorous confirmation of the expectation that most systems are non-integrable.

Notably, the integrable systems naturally emerge as exceptional cases in the proof of non-integrability without resorting any existing techniques for integrable systems, that is, the Bethe ansatz and the Jordan-Wigner transformation to free fermions.
Therefore, the non-integrability proof method is expected to help the discovery of novel integrable systems without relying specific tools for integrable systems.
Also, the result in the companion paper \cite{FULLPAPER}, along with Refs.~\cite{nozawa2020explicit, fukai2023all,chiba2024proof}, shows that expansion of physical quantities and coefficient comparison is useful not only for proving non-integrability, but also for construction of local conserved quantities of both integrable and non-integrable systems.

In the class of models in consideration, the presence or absence of $k$-local conserved quantities for $3\leq k \leq N/2$ follow the same fate,
i.e.,  if a system has (resp. does not have) 3-local conserved quantity, then this system has (does not have) $k$-local conserved quantities for any $4\leq k \leq N/2$.
A similar property is conjectured for more general classes of spin systems in Ref. \cite{grabowski1995integrability}, where for a broad class of spin chains with nearest-neighbor interactions, the presence or absence of $3$-support conserved quantities is conjectured to determine the presence or absence of general $k$-support conserved quantities with $k\geq 4$.
Clarifying whether this conjecture holds or does not hold in a more general class of systems is left as an important future problem.

\paragraph*{Acknowledgments---}
The authors are grateful to \mbox{HaRu K. Park}, \mbox{Atsuo Kuniba}, \mbox{Chihiro Matsui}, \mbox{Kazuhiko Minami}, \mbox{Masaya Kunimi}, \mbox{Hal Tasaki}, and \mbox{Hosho Katsura} for fruitful discussion.
M.Y. is supported by WINGS-FMSP.
N.S. and Y.C. are supported by JST ERATO Grant No.~JPMJER2302, Japan.
Y.C. is also supported by Japan Society for the Promotion of Science KAKENHI Grant No.~JP21J14313 and the Special Postdoctoral Researchers Program at RIKEN.

\bibliography{main}

\clearpage
\widetext
\appendix

\end{document}